\documentclass[proceedings,preprint]{rmaa}

\usepackage{paralist}

\usepackage{psfrag,color}

\SetYear{2008}
\SetConfTitle{Magnetics Fields in the Universe II: From Laboratory and Stars to the Primordial Structures}

\title{Star formation triggered by SNR impact into magnetized neutral clouds} 

\author{
  M. R. M. Le\~ao, \altaffilmark{1,5} 
  E. M. de Gouveia Dal Pino, \altaffilmark{1,6}
  D. Falceta-Gon\c{c}alves, \altaffilmark{2,3,7}
  C. Melioli, \altaffilmark{4,8}
  F. G. Geraissate \altaffilmark{1,9}}

\altaffiltext{1}{IAG, Universidade de S\~ao Paulo, Brazil.}
\altaffiltext{2}{Universidade Cruzeiro do Sul, Brasil.}
\altaffiltext{3}{University of Wisconsin-Madison, USA}
\altaffiltext{4}{University of Bologna, Italy.}
\altaffiltext{5}{mrmleao@astro.iag.usp.br}
\altaffiltext{6}{dalpino@astro.iag.usp.br}
\altaffiltext{7}{diego.goncalves@unicsul.br}
\altaffiltext{8}{cmelioli@astro.iag.usp.br}
\altaffiltext{9}{geraissate@astro.iag.usp.br}

\shortauthor{Le\~ao et al.}
\shorttitle{Star formation triggered by SNR}

\listofauthors{M. R. M. Le\~ao, E. M. de Gouveia Dal Pino, D. Falceta-Gon\c{c}alves, C. Melioli \& F. G. Geraissate}
\indexauthor{Le\~ao, M. R. M.}
\indexauthor{de Gouveia Dal Pino, E. M.}
\indexauthor{Falceta-Gon\c{c}alves, D.}
\indexauthor{Melioli, C.}
\indexauthor{Geraissate, F. G.}

\abstract{Present day star formation (SF) takes place in giant molecular clouds (GMCs). These contain a wealth of structures on all length-scales with highly supersonic motions and it is believed that these supersonic motions induce the observed density inhomogeneities in the gas that drive star formation. Suggested candidates for driving supersonic motions and SF include supernova shocks.  Considering the physical conditions that are relevant for triggering star formation in interactions involving SN shocks and neutral clouds, we have built diagrams of the SNR radius versus the cloud density in which the conditions above constrain a shaded zone where star formation induced by SN shock front-cloud interactions is allowed. The diagrams are also tested with fully 3-D MHD radiative cooling simulations involving a SNR and a self-gravitating cloud and we find that the numerical analysis is consistent with the results predicted by the  diagrams.  While the inclusion of a homogeneous magnetic field approximately perpendicular to the impact velocity of the SNR with an intensity $\sim 1 \; \mu$G within the cloud results only a small shrinking of the star formation zone in the diagrams, a larger magnetic field ($\sim 10\;\mu$G) causes a significant shrinking, as expected. Applications of our results to real star formation regions in our own galaxy have revealed that their formation could have been triggered by a shock wave produced by a SN explosion under specific values of the initial cloud density and the SNR radius. Finally, we have evaluated the effective global star formation efficiency  of this sort of interactions and found that it is smaller than the observed values in our own Galaxy (SFE $\sim$ 0.01-0.3). This result is consistent with previous  work in the literature and also suggests that the mechanism presently investigated, though very powerful to drive structure formation, supersonic turbulence and eventually, local star formation, does not seem to be sufficient to drive $global$ star formation in normal star forming galaxies, nor even when the magnetic field in the neutral clouds is neglected. }

\addkeyword{star: star formation}
\addkeyword{ISM: clouds}
\addkeyword{magnetic fields}
\addkeyword{supernova remnant}

\begin{document}
\maketitle

\section{Introduction}
\label{sec:intro}

Essentially all present day star formation takes place in molecular clouds (MCs; e.g. Blitz 1993; Williams, Blitz \& McKee 2000). It is likely that the MCs are relatively transient, dynamically evolving structures  produced by compressive motions in the diffuse HI medium of either gravitational or turbulent origin, or some combination of both (e.g., Hartmann et al. 2001; Ballestero-Paredes et al. 2006). In fact, observations of supersonic line-widths in the MCs support the presence of supersonic turbulence in these clouds with a wealth of structures on all length-scales (Larson 1981; Blitz \& Williams 1999; Elmegreen \& Scalo 2004; Lazarian 1999). Recent numerical simulations in periodic boxes have shed some light on the role of turbulence in the evolution of the MCs and star formation within them. They suggest that the $continuous$ injection of supersonic motions, maintained by internal or external driving mechanisms (see below), can support a cloud  $globally$ against gravitational collapse so that the net effect of turbulence seems to be  to inhibit collapse and this would explain the observed low overall star formation efficiencies in the Galaxy  (Klessen et al 2000; Mac Low \& Klessen 2004; Vazquez-Semandeni et al. 2005). On the other hand, the supersonic turbulence is also able to produce density enhancements in the gas  that may allow $local$ collapse into stars in both nonmagnetized (Klessen et al. 2000;  Elmegreen \& Scalo 2004) and magnetized media (Heitsch et al. 2001; Nakamura \& Li 2005).

Suggested candidates for an internal driving mechanism of turbulence include feedback from both low-mass and massive stars. These later, in particular, are major structuring agents in the ISM  in general (McCray \& Snow 1979),  initially through  the production of powerful winds and  intense ionizing radiation and, at the end of their lives through the  explosions as supernovae (SNe). It is worth noting however, that  MCs with and without star formation have similar kinematic properties (Williams et al. 2000). External candidates include galactic spiral shocks (Roberts 1969; Bonnell et al. 2006) and again SNe shocks (Wada \& Norman 2001; Elmegreen \& Scalo 2004). These processes seem to have sufficient energy to explain the kinematics of the ISM and can generate the observed velocity dispersion-sizescale relation (Kornreich \& Scalo 2000). Other mechanisms, such as magnetorotational instabilities, and even the expansion of HII regions and fluctuations in the ultraviolet (UV) field apparently inject energy into the ambient medium at a rate which is about an order of magnitude lower than the energy that is required to explain the random motions of the ISM at several scales, nonetheless the relative importance of all these injection mechanisms upon star formation is still a matter of debate (see, e.g., Joung \& MacLow 2006;  Ballesteros-Paredes et al. 2006; MacLow 2008 for reviews).

In this work, we focus on one of these driving mechanisms $-$ supernova explosions produce large blast waves that strike the interstellar clouds, compressing and sometimes destroying them, but the compression by the shock may also trigger local star formation (Elmegreen \& Lada 1977; see also Nakamura et al. 2005 and references therein). 

The collective effect of the SNe is likely to be the dominant contributor to the observed supersonic turbulence in the Galaxy (Norman \& Ferrara 1996; Mac Low \& Klessen 2004), however, it seems to inhibit $global$ star formation rather than triggering it (Joung \& MacLow 2006). Here, instead of examining the global effects of multiple SNe explosions upon the evolution of the ISM and the MCs, we will explore  the $local$ effects of these interactions. To this aim, we will consider a supernova remnant (SNR) either in its adiabatic or in its radiative phase impacting with an initially homogeneous diffuse neutral cloud and show that it is possible to derive analytically a set of conditions that can constrain a domain in the relevant parameter space where these interactions may lead to the formation of gravitationally unstable, collapsing structures. Besides, including the effects of the magnetic fields, we will apply this analysis to few SF regions with some indication of recent-past interactions with SN shock fronts (e.g., the Edge Cloud 2,  Yasui et al. 2006, Ruffle et al. 2007; and the "Great CO Shell", Reynoso \& Mangum 2001).  Finally, we will also test our analytically derived $SF$ $domain$ with 3D MHD simulations of SNR interactions with self-gravitating neutral clouds.

\section{SNR-Cloud Interactions and SF Constraints}
\label{sec:interaction}

A type II SN explosion generates a spherical shock wave that sweeps the interstellar medium (ISM), leading to the formation of a SNR. The interaction between a SNR and a cloud may compress the gas sufficiently to drive the collapse of the cloud. In order to describe analytically this interaction  we will consider a diffuse neutral cloud with initially homogeneous density and constant temperature. After the impact, an internal forward shock propagates into the cloud  with a velocity $v_{cs}$. In Melioli et al. (2006) and Le\~ao et al. (2008), we have derived the formulation for these interactions, taking into account both the effects of the curvature of the shock interaction and the magnetic field of the cloud. A set of constraints, which are described in the following paragraphs, determine the conditions for driving gravitational instability in the cloud right after the interaction with a SNR.

\subsection{The Jeans Mass Constraint}
\label{Jeans}

A first constraint gives the Jeans mass limit for the compressed cloud material. In the absense of magnetic fields and considering the interaction with a SNR in the adiabatic phase we find that

\begin{equation}
m_{J,a}\sim 750 \frac{T_{c,100}^2\;R_{SNR,50}^{1.5}}{I_5\;E_{51}^{0.5}} \;\; M_{\odot}
\label{eq:mjeansad}
\end{equation}

\noindent where $n_{c,10}$ is the cloud density in units of 10 $\rm cm^{-3}$, $E_{51}$ is the SN energy in units of $10^{51}\;\rm erg$, $R_{SNR,50}$ is the SNR shell radius in units of 50 pc and $I_5$ is a correction due to shock curvature (Le\~ao et al. 2008))

In terms of the SNR radius, the condition above may be expressed as:

\begin{equation}
R_{SNR,a}\lesssim 55.8\;\frac{E_{51}^{1/3}\;I_5^{2/3}\; n_{c,10}^{2/3}\;r_{c,10}^{2}}{T_{c,100}^{4/3}}\;\;\;pc
\label{eq:parametro1}
\end{equation}

When including the magnetic field in the cloud, the corresponding minimum (Jeans) mass that the shocked material must have in order to suffer gravitational collapse, in terms of the cloud pre-shock and the SNR parameters, is given by

\begin{equation}
m_{J,B} \simeq \frac{2100}{(y\;n_{c,10})^{1/2}}\left[\frac{4y B_{c,6}^2}{n_{c,10}} + 4.14  T_{c,100}\right]^{3/2} \;\; M_{\odot}
 \label{eq:parametro1mag}
\end{equation}

\noindent  where $B_{c,6}$ is the magnetic field of the cloud in the pre-shock in units of $10^{-6}\; G$.

We can obtain $R_{SNR}$ as function of $n_{c}$ and then obtain an approximate condition for gravitational collapse solving $m_c\geq m_{J,B}$, where $m_c$ is the cloud mass, and

 \[y=\frac{4}{2M^{-2} + M_A^{-2} + [(2M^{-2} + M_A^{-2})^2 + 8 M_A^{-2}]^{1/2}}\; .\]

We notice that in the limit that $\frac{B^2}{8\pi}\ll \rho\;c_s^2$ the equation above (\ref{eq:parametro1mag}) recovers the solution of Eq. (\ref{eq:mjeansad}).

\subsection{Constraint for non-destruction of the cloud due to a SNR impact}
\label{destr}

A second condition establishes the constraint upon the shock front under which it will $not$ be too strong  to destroy the cloud completely making the gas to disperse in the interstellar medium before becoming gravitationally unstable. In order to check this condition, a gravitationally unstable mode (with  typical time $t_{un}$) must grow in a time scale smaller than the cloud destruction time. This implies the following constraint for an interaction without magnetic field

\begin{equation}
R_{SNR,a} \gtrsim 52 \ {{E_{51}^{0.33} \ T_{c,100}^{0.44} \ I_5} \over{n_{c,10} \ r_{c,10}^{1.56}}} \ \ \ \ {\rm pc}
\label{eq:parametro2}
\end{equation}

\noindent with a SNR in the adiabatic regime.

When including the magnetic field in the cloud, we obtain the following condition for the Mach number of the shock

\begin{eqnarray}
\frac{M (1 + y \beta/3)^{3/8}}{y^{5/8}}\lesssim \; 18.7 \left(\frac{n_{c,10}}{T_{c,100}}\right)^{7/8}r_{c,10}^{7/4}
\label{eq:parametro2mag}
\end{eqnarray}

\noindent where $\beta=B_c^2/8\pi \rho_c c_s^2$, where $c_s$ is the sound speed in the cloud gas.

\noindent Substituting the relations  for $y$ and $M$ for a collision with an adiabatic SNR (see Le\~ao et al. 2008 for details) into the equation above we find numerically the new constraint over $R_{SNR,a}$ in order to not destroy the magnetized cloud at the impact and allow its gravitational collapse (see below).

\subsection{Penetration extent of the SNR shock front into the cloud}

Besides the constraints derived in the previous sections (\ref{Jeans} and \ref{destr}), a third conditions establishes the penetration extent of the SNR forward shock front inside the cloud before being stalled due to radiative losses. The shock must have energy enough  to compress as much cloud material as possible before fainting. 

In the absence of the magnetic field, we can show that this condition applied to an interaction involving an adiabatic SNR implies:

\begin{equation} 
    R_{SNR,a} \lesssim 170 {{E_{51}^{0.33} I_5^{0.66}} \over {(r_{c,10}\Lambda_{27})^{2/9} n_c^{0.5}}} \ \ \ {\rm pc}
 \label{eq:parametro3mag}
\end{equation}

\noindent where $\Lambda_{27}$ is the cooling function ($\Lambda$) in units of $10^{-27}$  erg cm$^{3}$ s$^{-1}$.

When considering the presence of the normal magnetic field in the cloud, our estimates result essentially the same constraint as in the absence of B (Eq. \ref{eq:parametro3mag}) for the typical fields observed in these clouds ($\sim 10^{-5} - 10^{-6}\;G$).

\subsection{Diagrams for SNR-Cloud interactions}
\label{diagrams}

The three constraints derived above in Sections 3.1, 3.2, and 3.3 for both non-magnetized and magnetized clouds interacting with SNRs in the adiabatic phase are plotted together in a diagram showing the SNR radius versus the initial (un-shocked) cloud density for different values of the cloud radius. 

Figure 1 compares these diagrams for non-magnetized and magnetized clouds. The three constraints establish a shaded zone in the parameter space of the diagrams where gravitational collapse of the shocked cloud material can occur. Only cloud$-$SNR interactions with initial physical conditions ($r_c$, $n_c$ and $R_{SNR}$) lying within the shaded region  may lead to a process of star formation.

We notice that the presence of a normal magnetic field to the shock front with an intensity of 1 $\mu$ G inhibits slightly the domain of SF in the diagrams (the dark shaded zone), as expected. The magnetic field plays a dominant role over the Jeans constraint that cause a drift of the allowed zone of SF (dark shaded zone) to the right in the diagrams (i.e., to higher cloud densities) when compared to the case without magnetic fields (the light shaded zone). When larger intensities of magnetic fields are considered (5-10 $\mu$ G) there is a significant shrinking of the allowed SF zone in the diagrams.

\begin{figure*}[!t]
  \makebox[0pt][l]{\textbf{a}}%
  \hspace*{\columnwidth}\hspace*{\columnsep}%
  \textbf{b}\\[-0.7\baselineskip]
  \parbox[t]{\textwidth}{%
     \vspace{0pt}
     \includegraphics[width=\columnwidth]{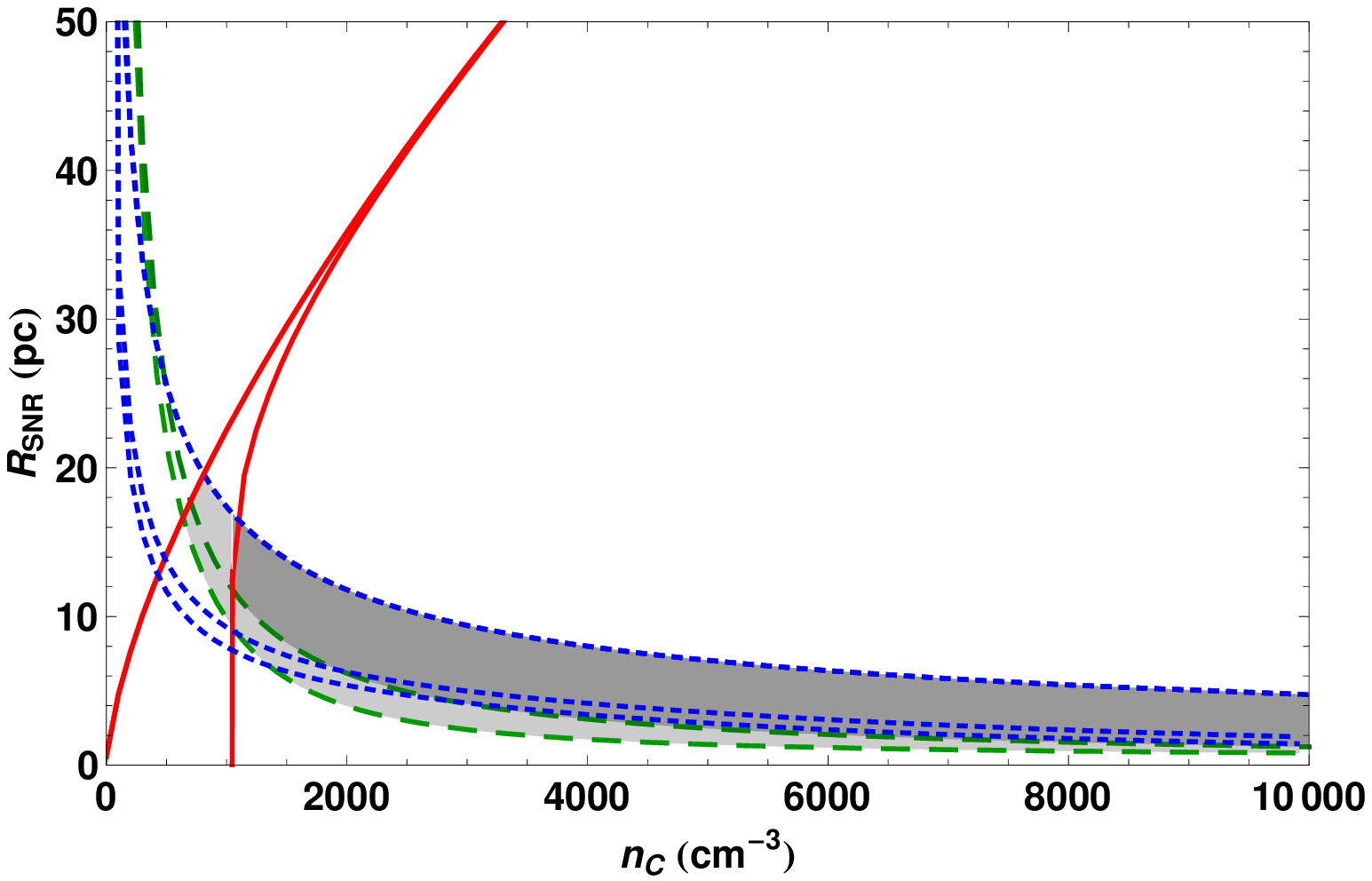}%
     \hfill%
     \includegraphics[width=\columnwidth]{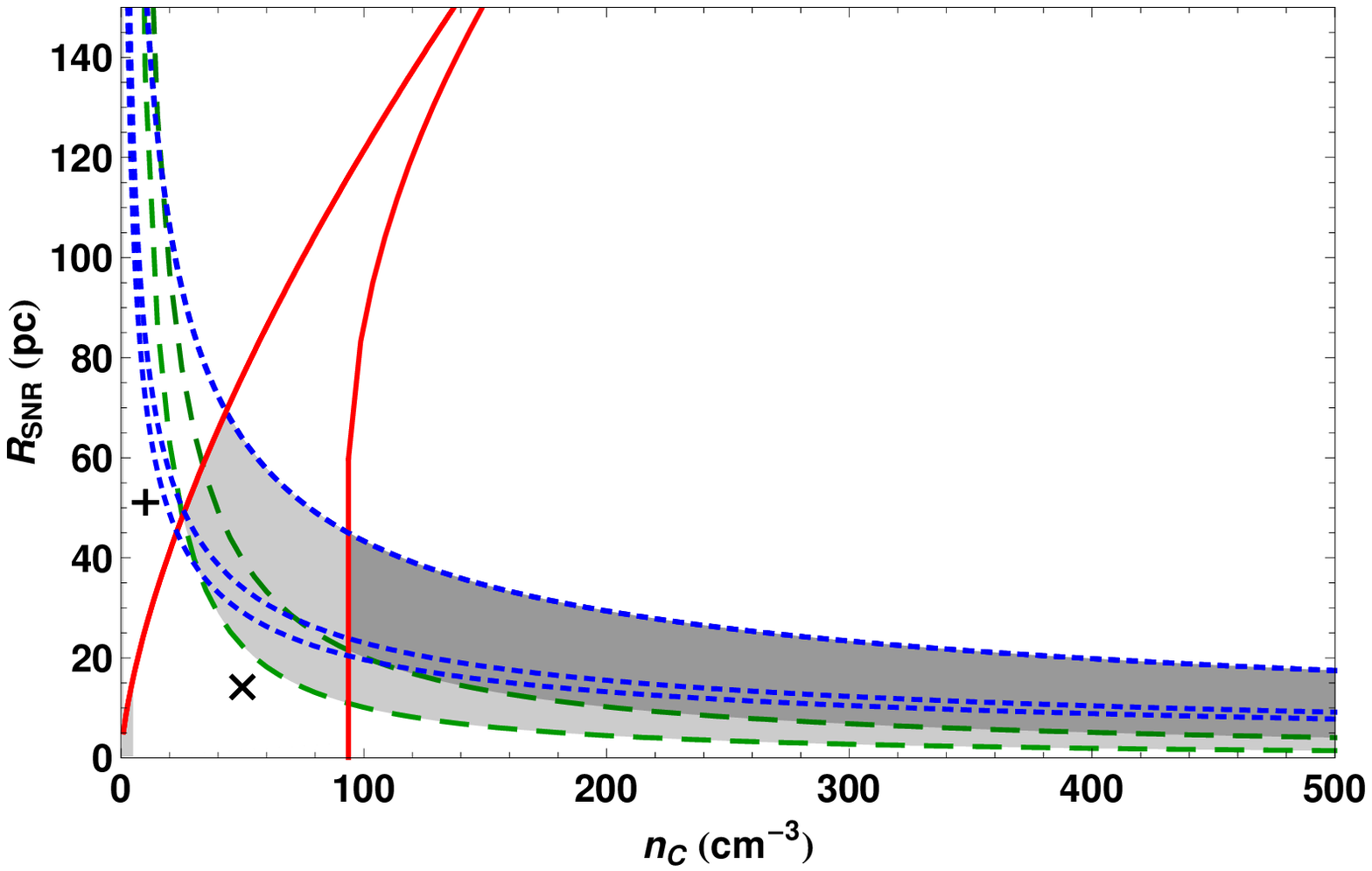}
     }
     \makebox[0pt][l]{\textbf{c}}%
  \hspace*{\columnwidth}\hspace*{\columnsep}%
  \textbf{d}\\[-0.7\baselineskip]
  \parbox[t]{\textwidth}{%
     \vspace{0pt}
     \includegraphics[width=\columnwidth]{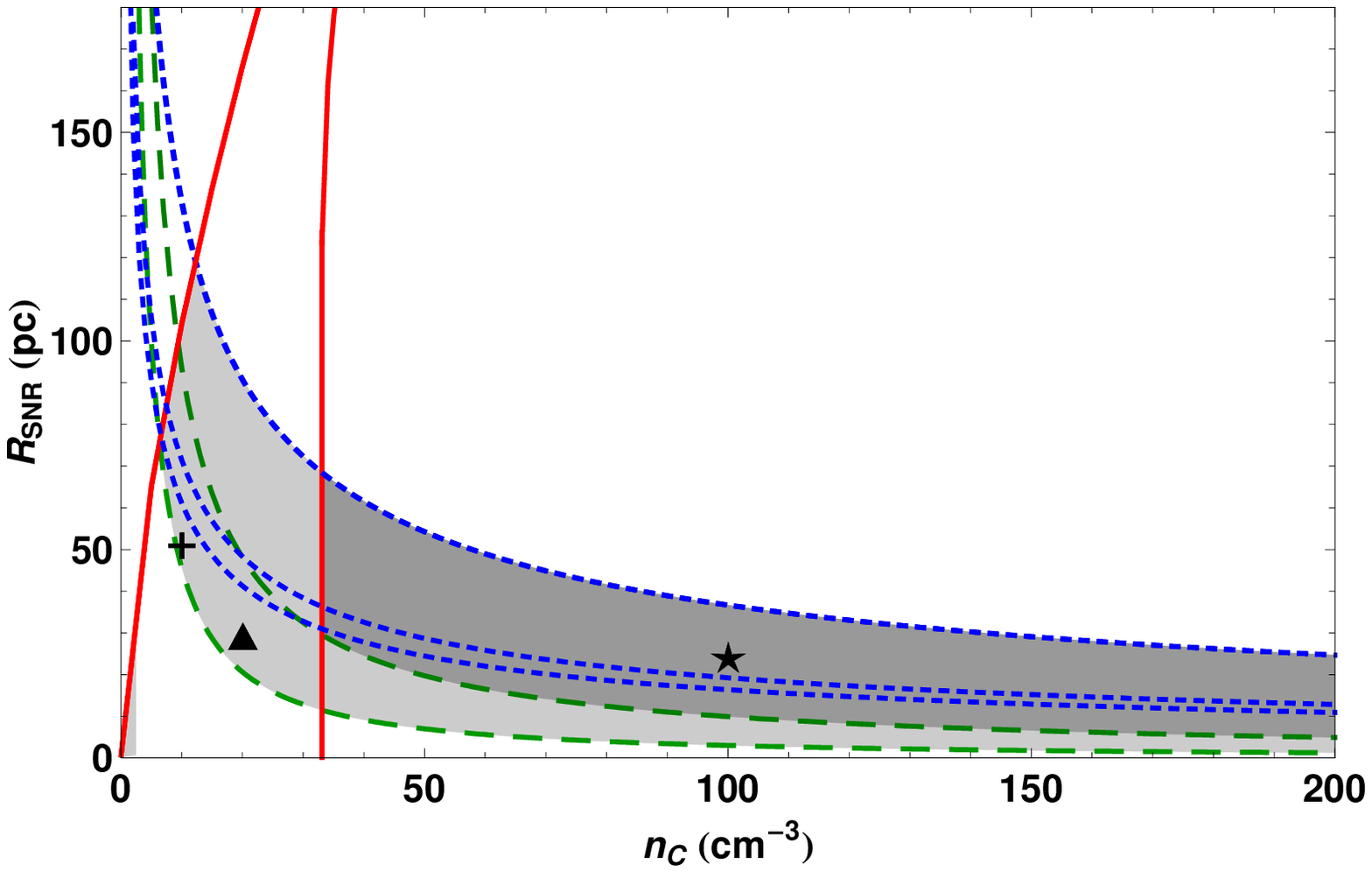}%
     \hfill%
     \includegraphics[width=\columnwidth]{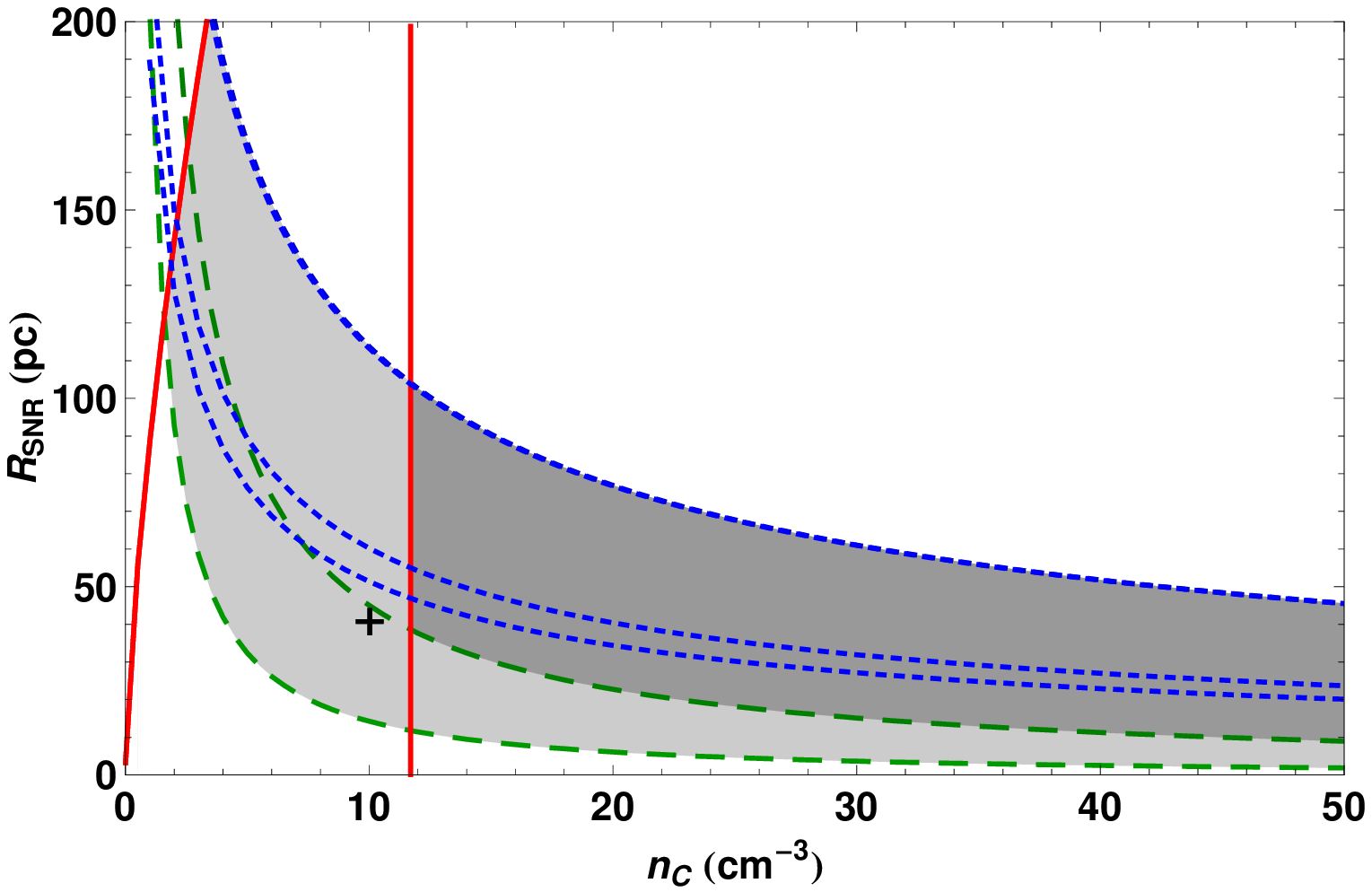}
     }
  \caption{Constraints on the SNR radius  versus cloud density for 4 different  cloud radius. Left-top panel: $r_c$ = 1 pc; right-top: $r_c$ = 5 pc; left- bottom panel: $r_c$ = 10 pc; and right-bottom panel: $r_c$ = 20 pc. Dashed (green) line: upper limit for complete cloud destruction after an encounter with an adiabatic SNR derived from Eqs. \ref{eq:parametro2} and \ref{eq:parametro2mag}; solid (red) line: upper limit for the shocked cloud to reach the Jeans mass derived from Eqs. \ref{eq:parametro1} and \ref{eq:parametro1mag} for an interaction with an adiabatic SNR; dotted (blue) lines: upper limits for the shock front to travel into the cloud before being decelerated to subsonic velocities derived from Eq. \ref{eq:parametro3mag} for different values of the cooling function $\Lambda(T_{sh})=  10^{-25}$ erg cm$^{3}$ s$^{-1}$ (lower curve), $5 \times 10^{-26}$ erg cm$^{3}$ s$^{-1}$ (middle curve), and $3 \times 10^{-27}$ erg cm$^{3}$ s$^{-1}$ (upper curve). The shaded zones define the region where star formation can be induced by a SNR-cloud interaction. The light one defines the SF region domain for a cloud with $B_c=0$ while the dark shaded zone defines the SF region domain for a magnetized cloud with $B=1\;\mu$G. The symbols in the panels indicate the initial conditions assumed for the clouds in the numerical simulations described in Section \ref{simulations} (see text for details).}
  \label{fig:comparacao}
\end{figure*}

The crosses and the triangle in Fig. \ref{fig:comparacao} indicate the initial conditions assumed for the SNR-clouds interactions examined in the numerical simulations described in Melioli et al. (2006) for unmagnetized SNR-cloud interactions. We see that when the magnetic field is included, all these symbols lie outside the SF domain of the diagrams. This means that for the initial conditions corresponding to them, SF is unlikely to occur.

\subsection{SNR-cloud interaction: numerical simulations}
\label{simulations}

In order to check the predictions of our semi-analytic diagrams built for interactions involving SNR shocks and clouds, we have also performed 3-D radiative cooling, magnetohydrodynamical (MHD) simulations taking into account self-gravity, in order to follow the late evolution of the shocked material within a magnetized cloud and check whether it suffers gravitational collapse or not in consistence with our diagrams.

\begin{figure*}[!t]
  \makebox[0pt][l]{\textbf{a}}%
  \hspace*{\columnwidth}\hspace*{\columnsep}%
  \textbf{b}\\[-0.7\baselineskip]
  \parbox[t]{\textwidth}{%
     \vspace{0pt}
     \includegraphics[width=\columnwidth]{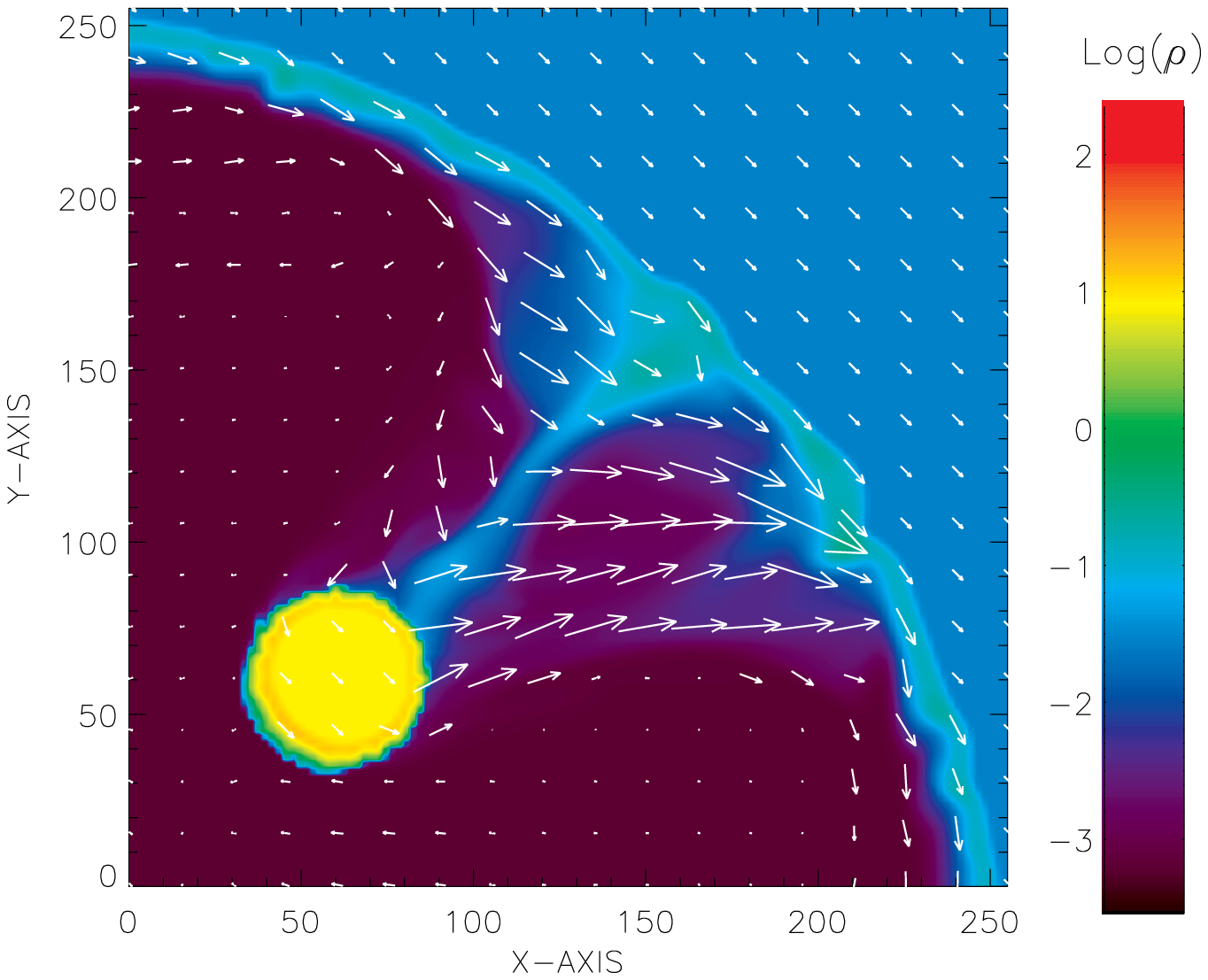}%
     \hfill%
     \includegraphics[width=\columnwidth]{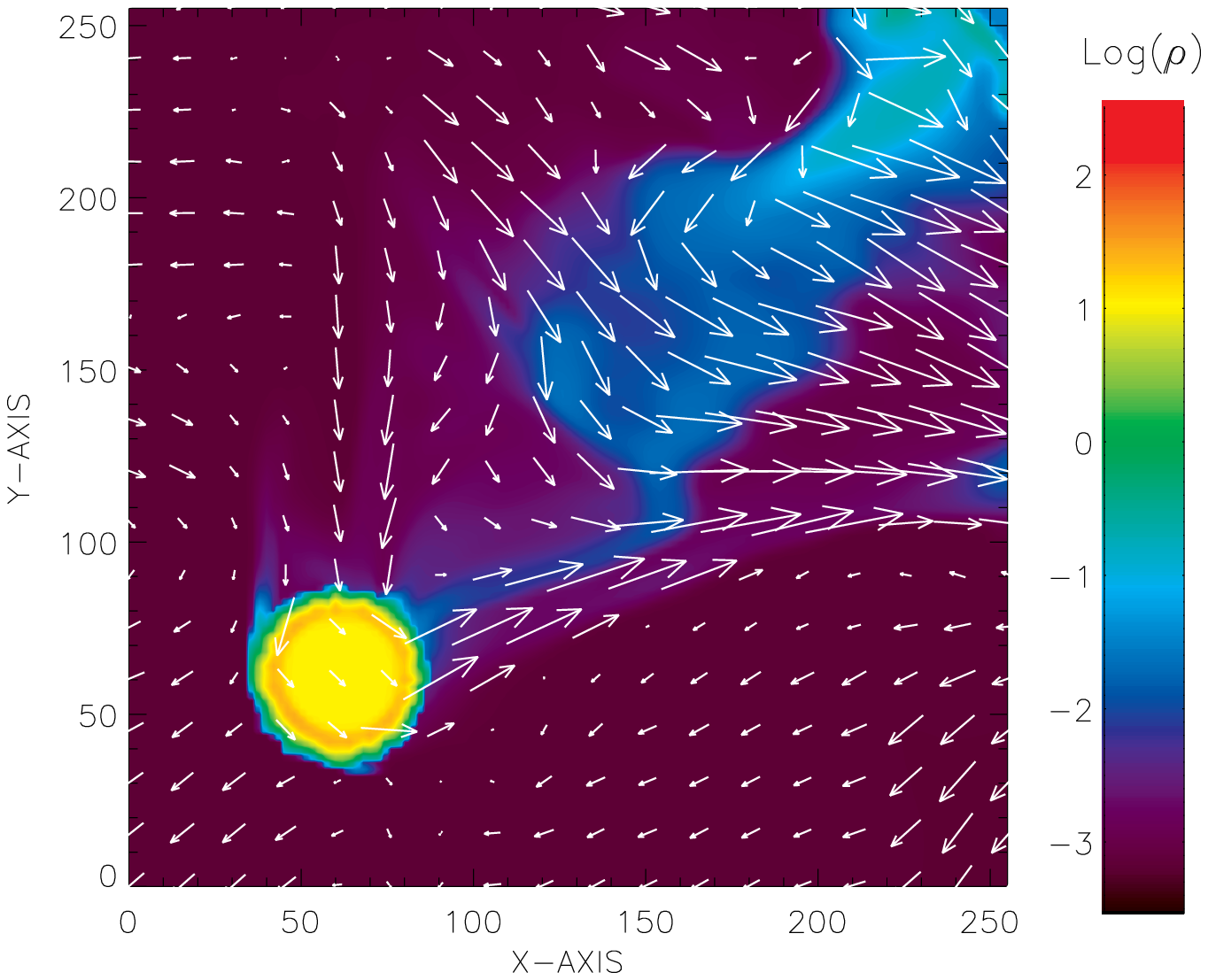}
     }
     \makebox[0pt][l]{\textbf{c}}%
  \hspace*{\columnwidth}\hspace*{\columnsep}%
  \textbf{d}\\[-0.7\baselineskip]
  \parbox[t]{\textwidth}{%
     \vspace{0pt}
     \includegraphics[width=\columnwidth]{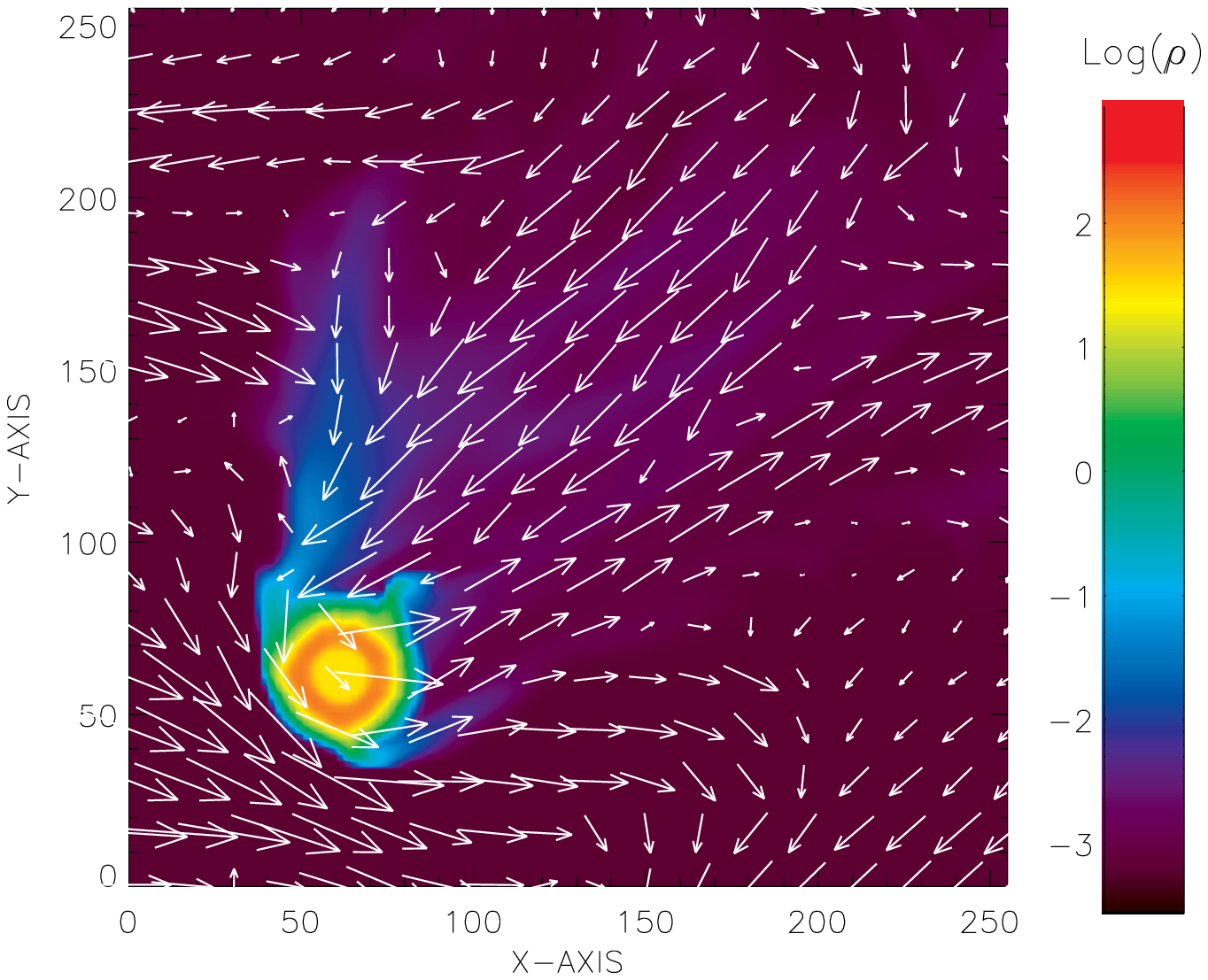}%
     \hfill%
     \includegraphics[width=\columnwidth]{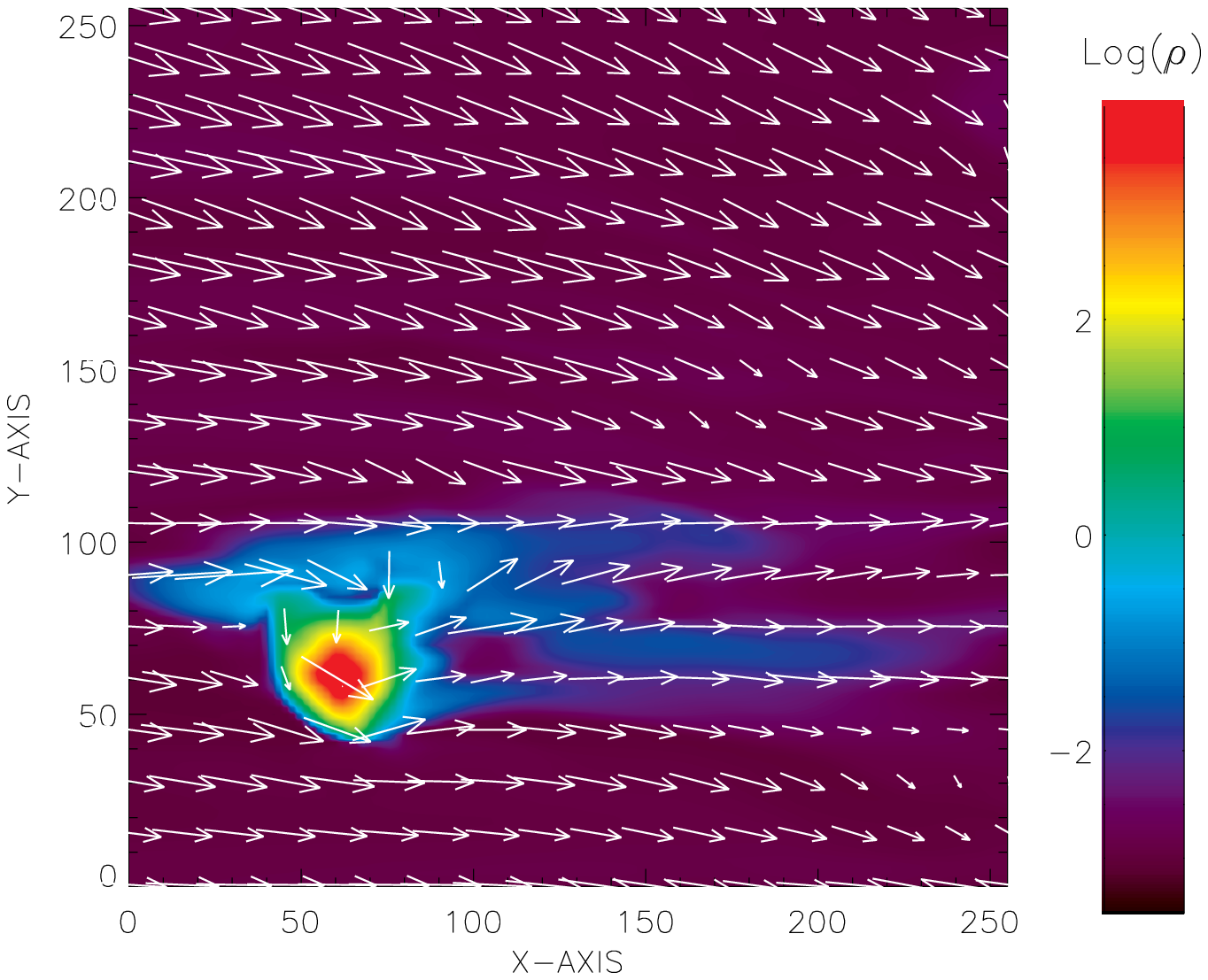}
     }
  \caption{Color-scale maps of the midplane density distribution (in log scale) evolution of the interaction between an expanding SNR and a magnetized cloud at a time a) $t=1.2 \times 10^6$ yr (top-left); b) $t=2. \times 10^6$ yr(top-right); c) $t=4 \times 10^6$ yr (bottom-left); and d)$t=5.6 \times 10^6$ yr (bottom-right). The SNR is generated by a SN explosion with an energy of $10^{51}$ erg. The ISM where the SNR expands has a number density $n=0.05$ cm$^{-3}$, and a temperature $10^4$ K. The cloud has an initial number density $n_c=100$ cm$^{-3}$, temperature $T_c=$ 100 K, radius $r_c=$ 10 pc and magnetic field $B_c=1\;\mu$G. The initial distance between the external surface of the cloud and the center of the SNR is $R_{SNR}=$ 25 pc. The arrows indicate the direction of B (extracted from Le\~ao et al. 2008).}
  \label{fig:simulacao}
\end{figure*}

As an example, Figure \ref{fig:simulacao} shows the evolution of a SNR$-$cloud interaction whose initial conditions correspond to the star labeled in the diagram of Figure \ref{fig:comparacao}c. According to the diagram these conditions would be able to generate a Jeans unstable core ($r_c$ = 10 pc, $n_c$ =100 cm$^{-3}$, $R_{SNR} = 25$ pc, $B_c=1\;\mu$G). The evolution of this system in the numerical simulation of Figure \ref{fig:simulacao} indicates that after $\sim$ 5.6 Myr a cold Jeans unstable core forms and collapses in agreement with the analytical results of Figure \ref{fig:comparacao}.

\section{Application to ISM}

We can apply the simple analytical study above to isolated star formation regions of our own galaxy. Here, we will address few examples that present some evidence of recent past interactions with SNRs.

\begin{figure}[!t]
   \includegraphics[width=\columnwidth]{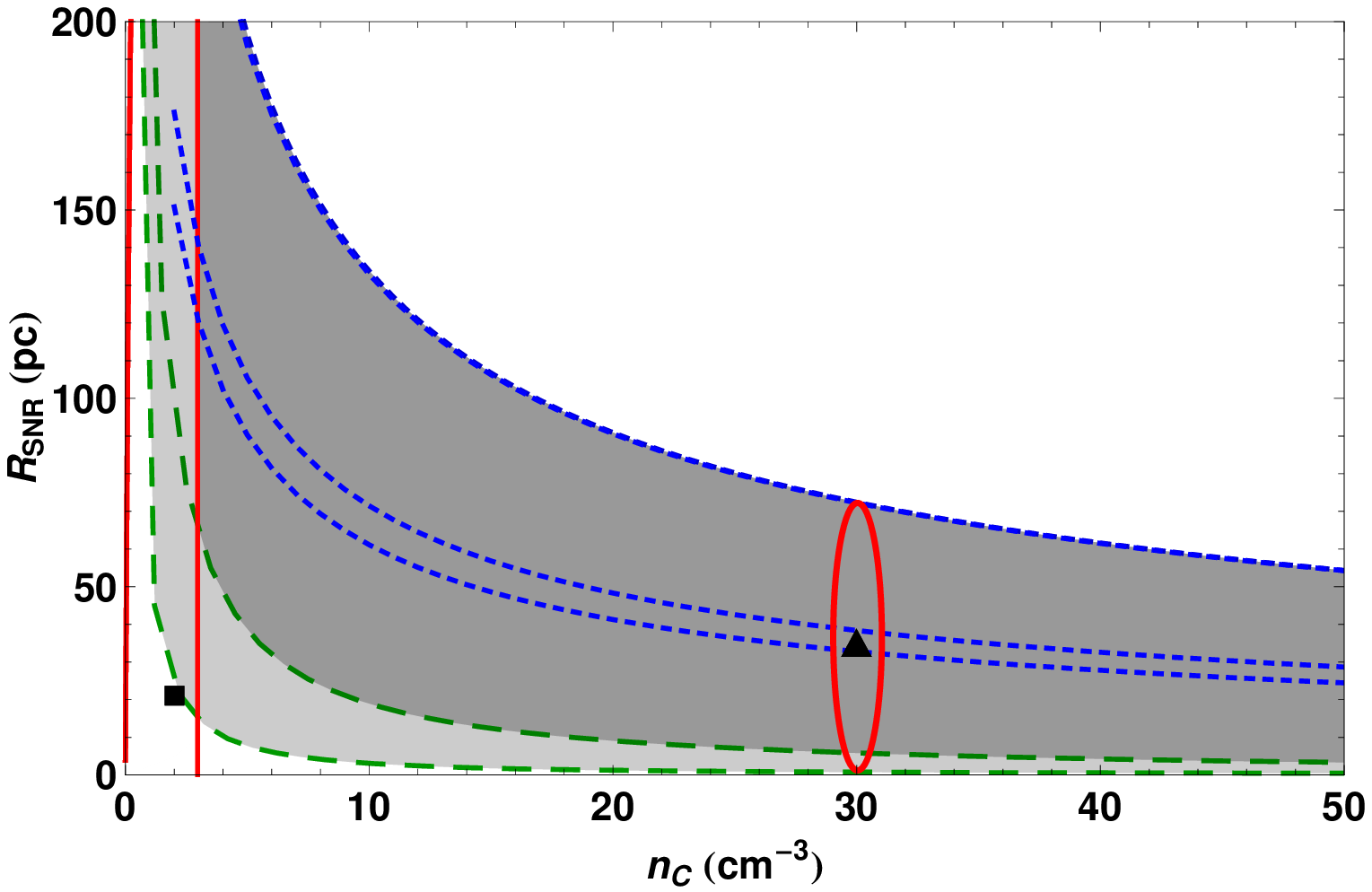}
   \caption{Diagram presenting the parameter space domain for SF for a cloud interaction with a SNR in the adiabatic regime. The thick lines correspond to the constraints for an interaction with an unmagnetized cloud while the light lines correspond to the constraints for an interaction with a magnetized cloud with  $B=1 \; \mu G$. As in the previous Figures, the dotted (blue) lines from top to bottom represent the shock penetration constraint for three different values of $\Lambda = 3\times 10^{-27} \;\rm erg\; cm^{3}\; s^{-1}$; $5\times 10^{-27} \;\rm erg\; cm^{3}\; s^{-1}$; and  $1\times 10^{-25} \; \rm erg\; cm^{3}\; s^{-1}$, respectively (see Le\~ao et al. 2008 for details). The other initial conditions are $r_c=$ 50 pc and $T_c=100\;\rm K$. The dark-gray shaded zone corresponds to the allowed SF zone for the interaction with magnetic field while the light-gray corresponds to the interaction without the magnetized cloud. The triangle represents the average conditions for the interaction involving the Large CO Shell while the (red) ellipse indicates a possible range of values for the interacting SNR radius. The square represents the initial conditions for Vela system (extracted from Le\~ao et al. 2008).} 
    \label{fig:compdiag50pcpts}
\end{figure}

The Large CO shell in the direction of Cassiopeia is an expanding structure at a velocity  $\sim 3\; km/s$ with a diameter of approximately  $100\; \rm pc$, a mass of $9.3 \times 10^{5}\; \rm M_{\odot}$, and a density of $\sim 35 \; \rm cm^{-3}$. Reynoso \& Mangum (2001) suggest that this expanding structure has probably originated from the explosion of a SN about $\sim 4 \times 10^6$ yr ago. Assuming that the cloud mass was originally uniformly distributed within a sphere of a radius of $\sim 50$ pc, the initial density would be  $n_c \simeq 30\; \rm cm^{-3}$. The SN shock front possibly induced the formation of the O9.5 star that has been detected as an IR source (IRAS 17146-3723). Presently, the Large CO Shell has an external radius of $50$ pc and an inner radius of $\sim 28$ pc (Reynoso \& Mangum, 2001). The age and small expansion velocity suggest that the SNR associated to the Large Shell system is presently a fainting evolved SNR. If we consider a cloud with the density above and a radius  $\sim $ 50 pc at the time of the potential interaction with a SNR in the adiabatic regime, we can identify this system in the SF diagram within the shaded zone, as indicated in Figure \ref{fig:compdiag50pcpts}, if the SNR had a radius  between $2.5 - 72$ pc. However, when we include a magnetic field in the cloud of $1\; \mu G$ the range of possible radii for the SNR is reduced to $R_{SNR} \sim 7.8 - 72$ pc if the maximum radius is calculated using a radiative cooling function  $\Lambda = 3\times 10^{-27}$ erg cm$^{3}$ s$^{-1}$.

The Edge Cloud 2 in the direction of Scorpius (Ruffel et al. (2007)) is another example of successful SF region that lies inside the shaded zones in our diagrams (see Le\~ao et al. 2008).

The SNR Vela, on the other hand, is a counter example. It has an almost spherical, thin HI shell expanding at a velocity of  $\sim 30\; km/s$. Instead of being impinging an interstellar cloud, it is expanding in a fairly dense environment with evidence of some structure formation. Assuming that Vela is at a distance $\sim 350$ pc from the Sun, its shell radius is of the order of 22 pc. The ambient density is $\sim$ 1 to 2 $\rm cm^{-3}$ and the initial energy of the SN was around  $1 - 2.5\times 10^{51}\;\rm erg$ (Dubner et al. 1998). These initial conditions correspond to the square symbol in the diagram of Figure \ref{fig:compdiag50pcpts} and it lies outside the SF shaded zone. This is consistent with the absence of dense clouds, clumps, filaments, or new born stars in the neighborhood of this SNR.

\section{Estimating the star formation efficiency}

In the study we have carried out here, we focused on isolated interactions between diffuse clouds and  SNRs without focusing on the  effects that such interactions can have upon the global SF in the Galaxy. The present star formation efficiency is typically observed to be very small, of the order of  few tens of percent in dispersed regions, but it can attain a  maximum of $\sim 0.3$ in cluster-forming regions (Lada \& Lada 2003; see also Nakamura \& Li 2006 for a review). 

We can try to estimate the star formation efficiency that interactions between SNRs and diffuse clouds produce and compare with the observed values in order to see the contribution of this mechanism upon the overall sfe in the Galaxy. In order to evaluate the corresponding global sfe of these interactions we have to estimate first their probability of occurrence in the Galaxy, 

\[f_{SNR-c}(R_{SNR})\simeq N_{SNII} \tau_{SNR}(R_{SNR}) \frac{A_{SNR}(R_{SNR})}{A_G}f_c\] where we have assumed a homogeneous galactic thin disk with a radius of 20 kpc to compute the galactic area, $A_G$, and where $N_{SNII}$ is the rate of SNII explosions (e.g. Cappellaro, Evans \& Turatto 1999), $ \tau_{SNR}(R_{SNR})$ is the lifetime of a SNR (Melioli et al. 2006, McCray 1985), $A_{SNR}(R_{SNR})$ is the SNR area and $f_c$ is the volume filling factor of the cold gas in the ISM $f_c \simeq 5 \%$ (e.g., de Avillez \& Breitschwerdt 2005).

Now, if we multiply this probability by the mass fraction of the shocked gas that is gravitationally unstable within the SF domain of our SNR$-$cloud interaction diagrams, we obtain an effective global star formation efficiency for these interactions:

\[sfe_{SNR-c}(R_{SNR})  \simeq f_{SNR-c}(R_{SNR}) \frac{m_{J}(R_{SNR})}{m_{c}}\]

\begin{figure*}[!t]
  \includegraphics[width=\columnwidth]{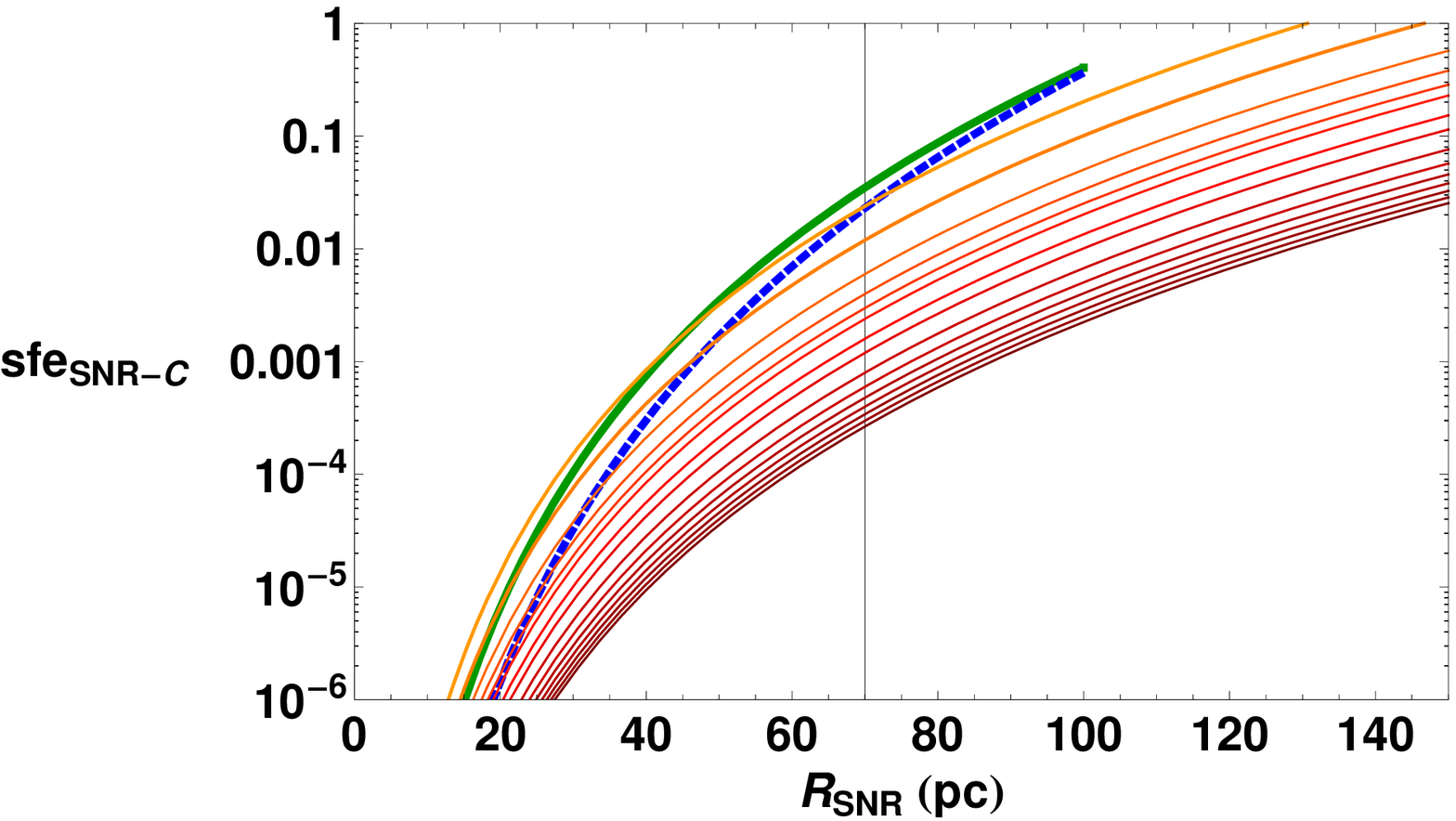}%
  \hspace*{\columnsep}%
  \includegraphics[width=\columnwidth]{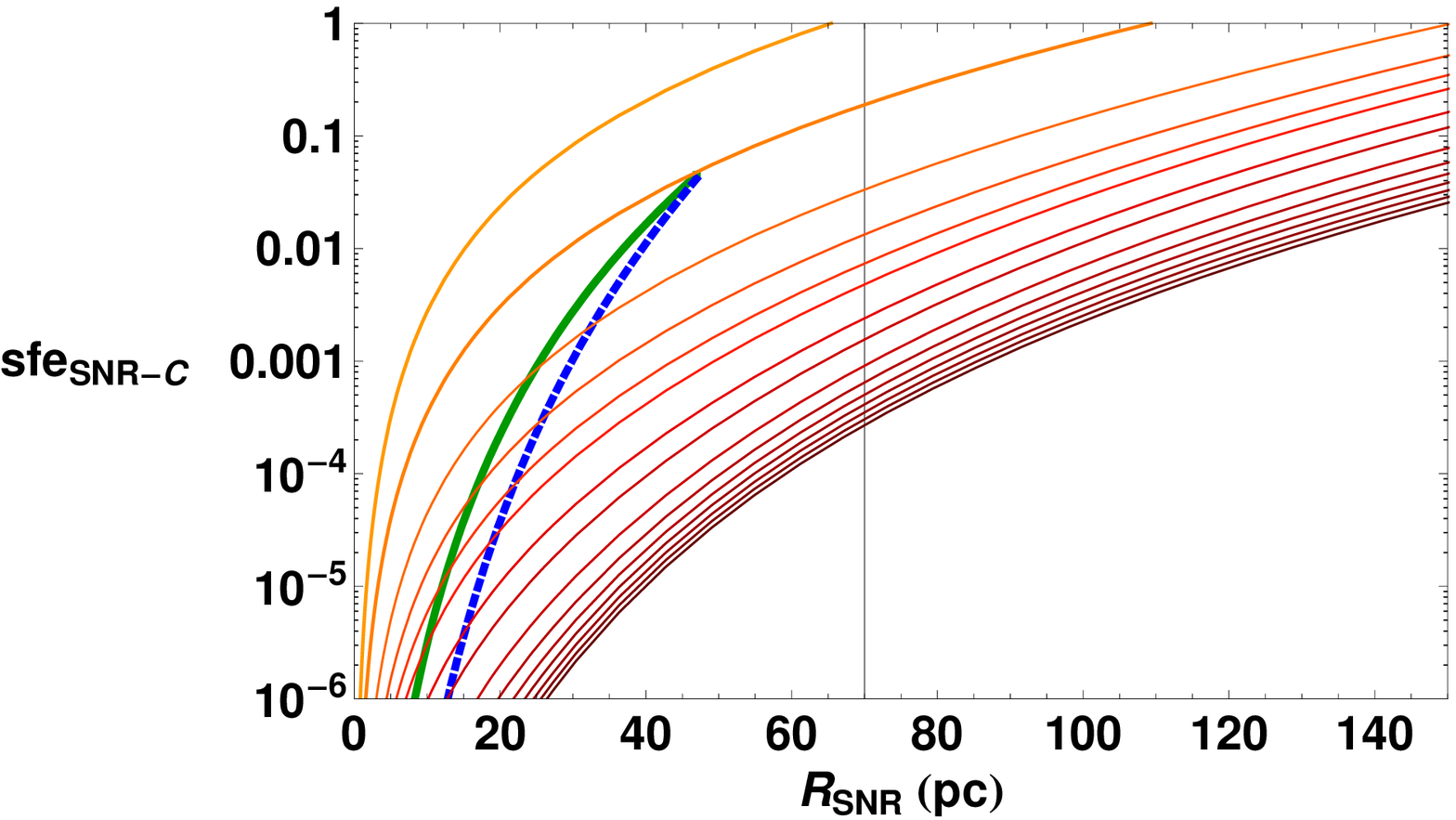}
  \caption{Plots of the calculated star formation efficiency as a function of the SNR radius in the adiabatic regime for several values of the cloud density represented by the solid (red) lines (from top to bottom: 10, 20, 40, 60, 80, 100, 150, 200, 300, 400, 500, 600, 700, 800, 900 $\rm cm^{-3}$) for an interaction with: (a) a cloud with $B_c=0$ (left panel); a cloud with $B_c=1\;\mu G$ (right panel). The dotted blue line represents the shock penetration condition and the solid green line the cloud non-destruction condition. Both constrain the allowed SF zone in these diagrams (extracted from Le\~ao et al. 2008).}
  \label{fig:sfe10pc}
\end{figure*}

\noindent As an example, Figure \ref{fig:sfe10pc} shows plots of the approximate sfe computed as a function of the SNR radius for a cloud with $B=0$ and a cloud with $B=1\;\mu$G and different values of the cloud density. The dotted and dashed lines in this figure represent the shock penetration and the cloud non-destruction conditions, respectively, examined before in this work and they constrain the allowed SF zone in these plots, as in the previous diagrams. The increase of the magnetic field tends to shift the SF zone to smaller values of the radius of the SNR both for SNR in the adiabatic and in the radiative regimes.

We note that the evaluated sfe for these interactions is smaller than the typical values observed for the Galaxy. This suggests that these powerful interactions are not sufficient to explain the observed sfe of the Galaxy either in the presence or in the absence of the magnetic field in the cloud. This result is consistent with previous analysis performed by Joung \& MacLow (2006) where these authors have concluded that Supernova-driven turbulence tends to inhibit global star formation rather than triggering it. We should note however, that they have based their conclusion on the computation of the star formation rate (SFR), rather than the sfe, from box simulations of the ISM with SN turbulence injection and their computed SFR has been weighed by a fixed value of the sfe taken from the observations (sfe $\sim$ 0.3).

\section{Conclusions}

We have presented here a study  of isolated interactions between SNRs and diffuse neutral clouds focusing on the determination of the  conditions that these interactions must satisfy in order to lead to gravitational collapse of the shocked cloud material and to star formation, rather than to cloud destruction. We have then built diagrams of the radius of the SNR as a function of the cloud density where this set of constraints delineate a domain within which star formation may result from these SNR-cloud interactions (Section \ref{diagrams}). As expected, we find that an embedded magnetic field in the cloud normal to the shock front with an intensity of 1 $\mu$ G inhibits slightly the domain of SF in the diagram when compared to the non-magnetized case.  

When larger intensities of magnetic fields are considered (5-10 $\mu$ G), the shrinking of the allowed SF zone in the diagrams is much more significant. We must emphasize however that, though observations indicate typical values of $B_c \simeq 5-10 \mu$ G for these neutral clouds, the fact that we have assumed uniform, normal fields in the interactions have maximized their effects against gravitational collapse. We should thus consider as  more realistic the result obtained when an effective $B_c \simeq 1 \mu$ G was employed. These diagrams derived from simple analytical considerations provide a useful tool for identifying sites where star formation could be triggered by the impact of a SN blast wave. 

We have also performed fully 3D radiative cooling MHD numerical simulations of the impact between a SNR and a self-gravitating cloud for different initial conditions (in Section \ref{simulations}) tracking the evolution of these interactions. We have found  the numerical results to be consistent with those established by the SNR$$cloud density diagrams.

We have applied the results above to a few examples of regions in the ISM with some evidence of  interactions of the sort examined in this work. In the case of the expanding Great CO Shell$-$O9.5 star system, we find that local star formation could have been induced in this region if, at the time of the interaction, the SNR that probably originated this expanding shell was still in the adiabatic phase and had a radius between $\sim$ 8 pc $-$ 29 pc, and impinged a magnetized cloud with  density around 30 $ \rm cm^{-3}$ (Figure \ref{fig:compdiag50pcpts}). Another example is the SF region near the Edge Cloud 2. This is one of the most distant cloud complexes from the galactic center where external perturbations should thus  be rare. But the recent detection of two young associations of T-Tauri stars in this region could have been formed from the interaction of a SNR in the radiative phase with a cloud,  if the interaction started $\lesssim 10^6$ yr, and the SNR had a radius $R_{SNR} \simeq$ $46$ pc - $84$ pc and the magnetized cloud a density around $n_c\sim14\; \rm cm^{-3}$ (see Le\~ao et al. 2008 for details).

Finally, though in this study we have focused on isolated interactions involving SNRs and clouds, we  used the results of the diagrams to  estimate the contribution of these interactions to global star formation. Our evaluated effective star formation efficiency for this sort of interactions is generally smaller than the observed values in our own Galaxy (sfe $\sim$ 0.01-0.3) (Figure \ref{fig:sfe10pc}). This result seems to be consistent  with previous analysis (e.g., Joung \& MacLow 2006) and suggests that these interactions are powerful enough to drive structure formation, supersonic turbulence (see, e.g., simulation of Figure \ref{fig:simulacao}) and eventually local star formation, but they do not seem to be sufficient to drive $global$ star formation in our galaxy or in other normal star forming galaxies, nor even when the magnetic field in the neutral cloud is neglected.

\end{document}